\documentclass[%
reprint,
superscriptaddress,
showpacs,
amsmath,amssymb,
aps,
prx,
longbibliography,
]{revtex4-1}

\usepackage{psfrag,graphicx,epsfig,color}% Include figure files
\usepackage{dcolumn}% Align table columns on decimal point
\usepackage{bm}% bold math
\usepackage{natbib}
\usepackage[usenames,dvipsnames,svgnames,table]{xcolor}
\usepackage{subfigure}
\usepackage{rotating}

\def\re    {R_\lambda}
\def\uu {{\mathbf{u}}}

\def\AA {{\mathbf{A}}}
\def\HH {{\mathbf{H}}}
\def\FF {{\mathbf{F}}}
\def\SS {{\mathbf{S}}}
\def\RR {{\mathbf{R}}}
\def\TT {{\mathbf{T}}}
\def\BB {{\mathbf{B}}}
\def\II {{\mathbf{I}}}

\DeclareMathOperator{\tr}{Tr}
\definecolor{mygreen}{rgb}{0,0.7,0.}

\begin{document}

\title{
}

\title{
Forecasting small scale dynamics
of fluid turbulence using deep neural networks
}

%%% Authors %%%
\author{Dhawal Buaria }
\email[]{dhawal.buaria@nyu.edu}
%\thanks{}
\affiliation{Tandon School of Engineering, New York University,
New York, NY 11201, USA}
\affiliation{Max Planck Institute for Dynamics and Self-Organization,
37077 G\"ottingen, Germany}
\author{Katepalli R. Sreenivasan}
\affiliation{Tandon School of Engineering, New York University,
New York, NY 11201, USA}
\affiliation{Department of Physics and the Courant Institute of Mathematical Sciences,
New York University, New York, NY 10012, USA}

\date{\today}% It is always \today, today,
             %  but any date may be explicitly specified

%\thispagestyle{empty}

\begin{abstract}
Turbulent flows consist of a wide range of interacting scales.
Since the scale range increases as some power of the flow Reynolds number, a faithful
simulation of the entire scale range is prohibitively expensive
at high Reynolds numbers. The most expensive aspect concerns the small scale motions;
thus, major emphasis is placed on understanding
and modeling them, taking advantage of their putative universality.
In this work, using physics-informed deep learning methods,
we present a modeling framework to capture
and predict the small scale dynamics of turbulence,
via the velocity gradient tensor.
The model is based on obtaining functional closures
for the pressure Hessian
and viscous Laplacian contributions
as functions of velocity gradient tensor.
This task is accomplished using deep neural networks
that are consistent with physical constraints
and incorporate Reynolds number
dependence explicitly
to account for small-scale intermittency.
We then utilize a massive
direct numerical simulation database,
spanning
two orders of magnitude in the large-scale Reynolds number,
for training and validation.
The model learns from low to moderate Reynolds numbers,
and successfully predicts velocity gradient
statistics at both seen and higher (unseen) Reynolds numbers.
The success of our present approach
demonstrates the viability of deep learning over traditional modeling
approaches in capturing and predicting small scale features of
turbulence.

\end{abstract}

\maketitle

\section{Introduction}

Turbulent fluid flows, 
ubiquitous in nature and technology, 
are characterized by strong and chaotic 
fluctuations across
a wide range of interacting scales in space and time. 
Such multiscale interactions are  highly nonlinear,
leading to mathematical intractability of the
governing equations. 
%are very challenging to characterize,
%but once mastered,
%may shed light on a number of related many-body problems. 
%There is a better hope of success in turbulence because the equations 
%of motion are known to adequately describe the physics.
Consequently, turbulence has defied an adequate 
framework despite a sustained effort in physics, mathematics 
and engineering, and our present understanding
remains incomplete, often relying on phenomenological approaches
\cite{MY.II, Frisch95, Sreeni97}.
An essential notion in this regard is that of small scale 
universality \cite{K41a}, which forms the backbone of turbulence theories and
models. It stipulates that, while the large scales are non-universal because 
of their dependence on flow geometry and 
energy injection mechanisms, such dependencies become progressively weaker as 
energy cascades to smaller scales,
ultimately endowing them with some form of universality that depends 
only on a few parameters of the flow.

From this perspective, universality requires sufficiently
large separation between the scales at which the energy is injected and those at 
which it is dissipated into molecular motion. This scale separation is determined
by the Reynolds number, $Re$ \cite{MY.II}; thus, investigating
universality requires data at high $Re$. 
While such high Reynolds numbers are attainable in some laboratory flows
and all geophysical flows, most quantities pertaining to small scales are still 
very difficult to measure  \cite{Wallace09}. 
Alternatively, direct numerical simulations (DNS) of the governing equations,
where the entire range of scales is resolved on a computational
mesh \cite{mm98}, 
provide information on every quantity desired. 
However, DNS is extremely expensive,
with recent studies showing that its
cost scales even faster than the traditional estimate of $Re^3$ 
\cite{YS:05, BPBY2019, BP2022}.
Thus, despite the rapid advances of high performance computing, 
high-$Re$ DNS, representative of natural
and engineering flows,  remains unlikely
for the foreseeable future.

Motivated by these considerations, 
we devise here an alternative approach based on
machine learning techniques to 
characterize the small scales of turbulence.
In recent years, the use of machine learning,
especially deep learning, has ushered in a new paradigm in 
various scientific fields \cite{lecun15}. 
The field of turbulence is no
different and there has been a flurry of machine learning methods
to improve turbulence modeling \cite{duraisamy19,Pandey}. 
A vast majority of them utilize the
framework of supervised learning \cite{good2016}, 
where neural networks are trained on input
data against labeled output data, although other paradigms have also be 
used \cite{mohan2019, kim2021, novati21}. 
The learning is also often `physics informed', i.e.,  
neural networks are  designed to satisfy some physical 
constraints, enabling efficient learning including 
significantly improved accuracy 
and stability \cite{karni21}. 
The approach utilized here follows a broadly similar paradigm, but
in a newly developed framework, specifically suited 
for small scales of turbulence. In particular, we
capture the small scale
dynamics of turbulence by training deep neural networks on existing 
DNS data at low and moderate $Re$, and demonstrate the
capability for predicting their dynamics at both seen $Re$ and higher unseen $Re$,
with important consequences for turbulence simulations. 

%low and moderate $Re$. Thereafter, we demonstrate its
%capability in predicting 
%the dynamics at higher unseen $Re$, as a possible alternative to 
%performing costly DNS in the future. 

The small scales of turbulence 
can be conveniently studied via the velocity
gradient tensor $\AA = \nabla \uu$,
where $\uu$ is the turbulent velocity field.
The tensor $\AA$ encodes various structural and statistical properties 
of turbulence, which are known to be universal to various degrees. 
The non-Gaussianity of its fluctuations
and the associated extreme events 
\cite{Siggia:81, zeff:2003, Ishihara07, schumacher2014small, BPBY2019},
the negative skewness of the
longitudinal (or diagonal) components associated with 
the energy cascade (from large to small scales) \cite{batchelor53, kerr85}, 
the preferential alignment of vorticity with the intermediate
strain eigenvector \cite{Ashurst87, BBP2020}, 
are a few notable examples.
By taking the gradient
of the incompressible Navier-Stokes equations, 
the evolution equation for $\AA$ becomes
\begin{align}
\frac{D \AA}{Dt} = -\AA^2 - \HH + \nu \nabla^2 \AA,
\label{eq:dadt}
\end{align}
where $H = \nabla \nabla P$ is the Hessian tensor
of the the kinematic pressure $P$ and 
$\nu$ the kinematic viscosity.
The above equation, studied by a number of past authors with different perspectives, 
dictates that
the velocity gradient tensor changes along a fluid element according to 
quadratic nonlinearity, the pressure effects
and viscous diffusion.
Since the trace $\tr(\AA)=0$ by incompressibility,
it follows that
\begin{align}
\nabla^2 P = \tr(\HH) = -\tr(\AA^2) \ . 
\label{eq:plap}    
\end{align}
That is, the pressure field is related to $\AA$ 
through a Poisson equation, implying that 
the pressure Hessian is non-local,
essentially coupling all scales of the flow.

In DNS, Eq.~\eqref{eq:dadt} is numerically solved 
on a large computational mesh by resolving all 
dynamically relevant scales \cite{mm98}.
However, an enticing approach is to develop a reduced-order
closure model for the pressure Hessian and the viscous Laplacian 
terms, written explicitly in terms of $\AA$, leading
to a fully local description \cite{Meneveau11}.
In this case, the dynamics of $\AA$ can be modeled
by an ordinary differential equation (ODE),
whereby statistical quantities of interest can be obtained, for example, 
by running Monte Carlo simulations of the ODE with 
arbitrary initial conditions.
Several such modeling attempts, see, e.g., Refs.
\cite{girimaji1990, CPS99, chevillard06prl, Wilczek2014Pressure, Lawson2015Velocity, Johnson2016Closure, Leppin2020Capturing},
have been made over the years,
including the use of neural networks more
recently \cite{parashar20, tian21}.
While these models
have enjoyed reasonable success, they still
struggle to capture various crucial aspects
of velocity gradient dynamics. In particular,
they have not been able
to capture the $Re$-dependencies of
velocity gradient statistics, which is 
a crucial aspect of small scale intermittency \cite{Frisch95,Sreeni97}.
To rectify such shortcomings is a primary motivation of the current work.
In the following section, we first provide the general framework
for developing the closure model and describe
the deep learning tools required for our purposes.

\section{Modeling using
tensor representation theory}

To obtain a functional closure,
the pressure Hessian and viscous Laplacian terms
in Eq.~\eqref{eq:dadt}
need to be specified as tensor functions of $\AA$. 
This can be most generally achieved 
by using tensor representation theory \cite{smith71, Zheng1994}, 
which has often been used in various modeling contexts 
\cite{pope1975, ling2016, parashar20, Leppin2020Capturing, tian21}.
Below, we briefly summarize the basic theory
and the novel changes introduced in this work 
to better capture the dynamics of $\AA$.

\subsection{Basic framework}

Tensor representation theory
allows us to express any desired (second order) tensor 
as a function of $\AA$. This is achieved 
by expressing the desired tensor
as a linear combination of tensors in an
appropriate tensor basis
constructed from $\AA$, 
with the coefficients that are functions 
of the scalar basis of $\AA$ \cite{smith71, Zheng1994}. 
To obtain the tensor and scalar bases 
the first step is to decompose $\AA$ into its 
symmetric and skew-symmetric parts, 
which are the strain-rate and rotation rate tensors, respectively:
\begin{align}
\SS = \frac{1}{2}(\AA + \AA^{\rm T}) \ , \ \ \ \RR = \frac{1}{2}(\AA - \AA^{\rm T}) \ .  
\end{align}
Using $\SS$ and $\RR$, we can construct
a general bases for tensors and scalars given 
in Eqs.~\eqref{eq:t10}-\eqref{eq:lam}.

%\begin{figure*}[bhp]
\begin{widetext}
\begin{align}
\begin{aligned}
&\TT^{(1)} = \SS \ , \ \ 
\TT^{(2)} = \SS \RR - \RR \SS  \ , \ \ 
\TT^{(3)} = \SS^2 - \frac{1}{3} \tr(\SS^2) \II \ , \ \
\TT^{(4)} = \RR^2 - \frac{1}{3} \tr(\RR^2) \II  \ , \ \
\TT^{(5)} = \RR \SS^2 - \SS^2 \RR  \ , \\
&\TT^{(6)} = \SS \RR^2 + \RR^2 \SS  - \frac{2}{3}  \tr(\SS \RR^2) \II  \ ,  \ \ 
\TT^{(7)} = \RR \SS \RR^2 - \RR^2 \SS \RR \ , 
\TT^{(8)} = \SS \RR \SS^2 - \SS^2 \RR \SS \ , \\
&\TT^{(9)} = \RR^2 \SS^2 + \SS^2 \RR^2 - \frac{2}{3} \tr(\SS^2 \RR^2) \II  \ , \ \ 
\TT^{(10)} = \RR \SS^2 \RR^2 - \RR^2  \SS^2 \RR  
\end{aligned}
\label{eq:t10}
\end{align}
\begin{align}
\begin{aligned}
&\BB^{(1)} = \RR \ , \ \ 
\BB^{(2)} = \SS \RR + \RR \SS \ , \ \ 
\BB^{(3)} = \SS^2 \RR + \RR \SS^2 \ , \ \ 
\BB^{(4)} = \RR^2 \SS - \SS \RR^2 \ , \\
&\BB^{(5)} = \RR^2 \SS^2 - \SS^2 \RR^2 \ , \ \ 
\BB^{(6)} = \SS \RR^2 \SS^2 - \SS^2 \RR^2 \SS  
\end{aligned}
\label{eq:b6}
\end{align}
\begin{align}
\begin{aligned}
\lambda_1 = \tr(\SS^2)  \ , \ \
\lambda_2 = \tr(\RR^2)  \ , \ \ 
\lambda_3 = \tr(\SS^3)  \ , \ \
\lambda_4 = \tr(\RR^2 \SS)  \ , \ \
\lambda_5 = \tr(\SS^2 \RR^2)  
\end{aligned}
\label{eq:lam}
\end{align}
\end{widetext}
%\end{figure*}

The ten $\TT^{(i)}$ in Eq.~\eqref{eq:t10} form the basis for symmetric tensors,
and the six $\BB^{(i)}$ in Eq.~\eqref{eq:b6} for skew-symmetric tensors;
$\lambda_i$ in Eq.~\eqref{eq:lam} is the basis of scalar invariants
required to determine the necessary coefficients. 
Since incompressibility gives $\tr(\SS)=0$
(and $\tr(\RR)=0$ trivially), it is 
easy to show
that $\tr(\TT^{(i)})=0$.
While the symmetric tensor basis $\TT^{(i)}$ 
is formulated to be trace-free owing to incompressibility,
a symmetric tensor basis does not in general
have to be trace-free; indeed such a basis will be somewhat different
from $\TT^{(i)}$ \cite{Zheng1994}.
Likewise, there are six scalar
invariants for a general tensor basis \cite{Zheng1994},
but incompressibility 
reduces the number to five, given in Eq.~\eqref{eq:lam} \cite{pope1975}. 

Using the above framework, we can
functionally model the pressure Hessian and viscous
Laplacian tensors.
However, note that while the
pressure Hessian tensor is symmetric, it
has a non-zero trace given by Eq.~\eqref{eq:plap}. 
Thus, we have to model the deviatoric part of the pressure Hessian 
tensor $\HH_{\rm d}$ as
\begin{align}
\HH_{\rm d} \equiv \HH - \frac{1}{3} \tr(\HH) \II 
 = \sum_{i=1}^{10} c_1^{(i)} (\lambda_1,...,\lambda_5) \TT^{(i)}, 
\label{eq:h10}
\end{align}
where the ten coefficients $c_1^{(i)}$ have to be determined
as functions of the scalar invariants.
The term $\tr(\HH)$ (and hence the isotropic
part of $\HH$) does not pose a closure problem,
since it can be written 
exactly in terms of $\AA$ as in Eq.~\eqref{eq:plap}.
The viscous Laplacian can be
decomposed simply into symmetric and skew-symmetric
contributions as
$\nabla^2\AA= \nu \nabla^2 \SS + \nu \nabla^2 \RR$,
which can be modeled using the respective tensor bases:
\begin{align}
 \begin{aligned}
\nabla^2 \AA 
= \sum_{i=1}^{10} c_2^{(i)} (\lambda_1,...,\lambda_5) \TT^{(i)} 
+ \sum_{i=1}^{6} c_3^{(i)} (\lambda_1,...,\lambda_5) \BB^{(i)}.
\end{aligned}
\label{eq:l16}
\end{align}
Here, we need to evaluate the 16 coefficients
$c_2^{(i)}$ and $c_3^{(i)}$
as functions of the scalar invariants.

%Note, tensor representation theory can in general 
%provide a basis for symmetric tensors which are not trace-free,
%which would be slightly different than 

This brief description nominally captures all previous attempts
to model the velocity gradient dynamics. 
For example, the pressure Hessian models developed in
\cite{Wilczek2014Pressure, Lawson2015Velocity, Johnson2016Closure, Leppin2020Capturing}
retain up to second order terms in Eq.~\eqref{eq:t10}, whereas the models
for viscous Laplacian retain (in the same works) just
the first order term. 
In Refs.~\cite{parashar20, tian21}, the pressure Hessian 
tensor was modeled in the same fashion as here. 
(However, no previous model accounted
for the $Re$-dependence of 
velocity gradient statistics, as we shall discuss in the next sub-section.)
In all these methods, the required scalar coefficients 
$c_1^{(i)}$ to $c_3^{(i)}$ were obtained as nonlinear 
scalar functions of $\AA$, 
satisfying a small set of the physical constraints.
However, one can more generally utilize the power
of deep learning to directly obtain the coefficients
$c_1^{(i)}$ to $c_3^{(i)}$, thus, in principle, learning all the
necessary physical constraints 
from the data itself \cite{parashar20, tian21}.

\subsection{Non-dimensionalization and Reynolds number dependence}

To utilize the above framework in conjunction with neural networks, 
it is important to non-dimensionalize all quantities.
There are several reasons.
Firstly, it is simply more convenient to work with
non-dimensional quantities in numerics. Secondly,
it facilitates efficient learning
(of network weights and biases), since the tensors in the bases are 
otherwise of different
orders in $\AA$. For instance, while the 
pressure Hessian is second order in $\AA$, 
the symmetric tensor basis spans
first to fifth order in $\AA$,
implying that the coefficients $c_1^{(i)}$
vary from order $1$ to $-3$ in $\AA$.
Appropriate non-dimensionalization renders
all coefficients to be of the same order, which leads
to better and faster learning \cite{good2016}.
Finally, non-dimensionalization also
allows us to appropriately introduce  
$Re$ as a parameter, whereby
the model system can be run at any chosen 
$Re$ to obtain desired 
(non-dimensional) statistics of velocity gradients. 

However, given the multiscale nature of turbulence,
the choice of variables for non-dimensionalization
is not unique. 
Since velocity gradients characterize small scales,
a natural choice is to utilize the Kolmogorov time and
length scales given as
\begin{align}
 \tau_K = ( \nu / \langle \epsilon \rangle )^{1/2}  \ , 
 \ \ \ \ \ \
 \eta_K = ( \nu^3 / \langle \epsilon \rangle )^{1/4}. 
 \label{eq:Kol}
\end{align}
Here, $\epsilon  = 2 \nu S_{ij} S_{ij}$ is the energy dissipation rate 
and $\langle \cdot \rangle$ denotes averaging
over space and time.
In homogeneous turbulence, we have $\langle S_{ij} S_{ij} \rangle  = \langle R_{ij} R_{ij} \rangle 
= \langle A_{ij} A_{ij} \rangle / 2 $. 
Thus, $\langle \epsilon \rangle = \nu \langle A_{ij} A_{ij} \rangle $,
giving $ \langle A_{ij} A_{ij} \rangle \tau_K^2 = 1$,
implying that $1/\tau_K$ quantifies the root-mean-square amplitude 
of $\AA$. 
This justifies the choice of using
Kolmogorov variables to non-dimensionalize $\AA$;
in fact, the above relations between the mean
quantities and $\tau_K$
allow us to impose convenient
constraints while running the model system
(as described in the Methods section). 
However, the choice is not obvious if one considers the extreme events
and hence higher-order statistics of $\AA$,
and also of pressure Hessian and viscous
Laplacian terms, since, due to intermittency,
they cannot be expected to scale on Kolmogorov variables \cite{YS:05, BPBY2019}.
We will persist with Kolmogorov normalization but introduce 
a phenomenological procedure to account for intermittency.

%are 
%also known to exhibit strong intermittency,
%meaning the extreme events cannot be described
%by $\tau_K$ \cite{BPBY2019}, which also must be accounted for
%(also discussed soon). 
%The choice of non-dimensionalization
%for the pressure Hessian and viscous
%Laplacian terms is less obvious. 
%As per the Kolmogorov 1941 theory, these terms 
%should also be prescribed by the Kolmogorov
%scales, but again due to multiscaling
%and intermittency they are not \cite{}.

In summary, the following non-dimensionalization is used:
$t^* = t/\tau_K$, $\mathbf{x}^* = \mathbf{x} /\eta_K$
(i.e., $\nabla^* =  \eta_K \nabla$), 
$\AA^* = \AA \tau_K$, $\HH^* = \HH \tau_K^2$ 
(and $\HH_{\rm d}^* = \HH_{\rm d} \tau_K^2$). We then obtain
the following equation for $\AA^*$:
\begin{align}
\frac{D \AA^* }{D t^*} = - ({\AA^*}^2 - \frac{1}{3} \tr({\AA^*}^2) \II ) 
- \HH_{\rm d}^* + {\nabla^*}^2 \AA^*  \ .
\label{eq:dadt2}
\end{align}
The terms $\HH_{\rm d}^*$ and ${\nabla^*}^2 \AA^*$
can now be modeled in terms of $\AA^*$ 
using the previously described framework, 
i.e., utilizing  Eq.~\eqref{eq:h10} and Eq.~\eqref{eq:l16}
where the tensor bases are appropriately
replaced by their non-dimensional counterparts,
i.e., $\TT^{*(i)}$ and $\BB^{*(i)}$,
and the coefficients $c_1^{(i)}$, $c_2^{(i)}$,
$c_3^{(i)}$ are dimensionless.
It can be seen immediately that the above
system does not have any Reynolds number dependence.
This is not surprising since 
we utilized Kolmogorov scales for non-dimensionalization,
temporarily choosing to ignore intermittency.
Thus, the Reynolds number dependence 
has to be reintroduced by hand (and validated {\em a posteriori}).  
Previous modeling attempts do not recognize this aspect and 
have consequently not captured the Reynolds number dependence;
see e.g., Refs.~\cite{Meneveau11, Johnson2016Closure, tian21}. 

%which at first glance does not appear correct.
%However, this is actually in line with the 
%K41 theory that
%all statistics can be universally described by
%Kolmogorov scales, i.e.,  non-dimensionalization
%by Kolmogorov scales would render all small-scale statistics
%independent of Reynolds number. 
%However, It is well known that the K41 theory
%fails due to intermittency, 
%leading to anomalous dependence of 
%velocity gradient statistics on Reynolds number;
%thus, the Reynolds number dependence 
%has to be reintroduced in an appropriate manner.
%Note, this is when the previous modeling
%attempts have also failed, or rather they only work
%to capture statistics 
%which do not depend on the Reynolds number,
%see e.g.
%\cite{Meneveau11, Johnson2016Closure, tian21}. 

It is worth stressing that there is no foolproof
way of introducing the $Re$-dependence into
the system. For instance, one could use
the large scales, say $L$ for length
and $U$ for velocity (and $L/U$ for time)
for non-dimensionalization, 
but it would be an incorrect choice for many aspects of the gradients. 
Instead, we devise the following pragmatic way
to introduce the Reynolds number dependence and intermittency effects. 
We maintain the non-dimensionalization
by Kolmogorov variables and
rescale the tensor bases as
\begin{align}
\label{eq:mph}
\HH_{\rm d}^* &= \sum_{i=1}^{10} c_1^{(i)} R_\lambda^{-\beta_1^{(i)}} \TT^{*(i)}  \ , \\
{\nabla^*}^2 \AA^* &= \sum_{i=1}^{10} c_2^{(i)} R_\lambda^{-\beta_3^{(i)}} \TT^{*(i)}  
+ \sum_{i=1}^{6} c_3^{(i)} R_\lambda^{-\beta_3^{(i)}} \BB^{*(i)} \ . 
\label{eq:mvl}
\end{align}
Here, $\re$ is the Reynolds number based on Taylor length scale 
(note that $\re \sim Re^{1/2}$ \cite{MY.II})
and the exponents $\beta_1^{(i)}$, $\beta_2^{(i)}$ and $\beta_3^{(i)}$
are additional model parameters which will be determined. This 
choice is motivated by two main reasons.
Firstly, the well-known multifractal description of 
turbulence suggests that velocity gradient statistics 
scale as power laws (or combinations of power laws) in $\re$ 
\cite{MY.II, Frisch95, Sreeni97}.
Secondly, the tensors in the bases span various orders of $\AA$, all of
which feel the intermittency effects differently.
Thus, the Reynolds number factors can rescale them
to the same order, allowing for more efficient learning,
essentially acting as additional physics-informed constraints to 
accommodate intermittency. 
If our physical understanding improves, it may well be possible to 
improve upon our present formulation.

\begin{figure*}[ht]
\begin{center}
\includegraphics[width=0.7\textwidth]{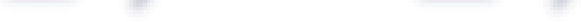}
\caption{
Reynolds number scaled tensor based neural network (ReS-TBNN) architecture 
utilized for modeling the deviatoric pressure Hessian,
based on Eq.~\eqref{eq:mph}. The first output layer and the second
input layer both have 10 nodes.
A similar network is utilized for the viscous Laplacian,
utilizing both the symmetric and skew-symmetric tensor bases,
as mentioned in Eq.~\eqref{eq:mvl}. The first output layer
and second input layer have both 16 nodes in this case.
}
\label{fig:tbnn}
\end{center}
\end{figure*}

\subsection{Reynolds-number-scaled tensor-based neural network (ReS-TBNN)}

We now consider the neural network architecture
utilized to model the unclosed terms. 
The tensor-based neural network (TBNN), utilizing only the symmetric 
basis from Eq.~\eqref{eq:t10}, was first proposed 
by Ling et al. \cite{ling2016} for turbulence modeling of the 
Reynolds stress tensor.
More recently, it was extended to modeling pressure Hessian in
\cite{parashar20, tian21}, 
while continuing to ignore Reynolds number effects. 
Unlike traditional neural networks, TBNN utilizes
two input layers. The architecture to model the pressure
Hessian is shown in Fig.~\ref{fig:tbnn}.
The first input layer uses the scalar basis $\lambda_i$,
which are then fed forward to multiple hidden layers to
obtain the scalar coefficients $c_1^{(i)}$ 
in the first output layer. 
The essence of this step is to model the scalar coefficients
as strongly nonlinear functions of the scalar basis -- 
an exercise traditionally performed by humans
\cite{Wilczek2014Pressure, Lawson2015Velocity, Johnson2016Closure}.
This is precisely the step where deep neural networks
are advantageous. 
The second input layer uses
the rescaled tensor basis as input, for instance 
$R_\lambda^{ -\beta_1^{(i)}} \TT^{(i)}$, 
for the pressure Hessian.
This second input layer
is contracted with the first output layer to obtain the
predicted pressure Hessian tensor in the final output layer,
in accordance with Eq.~\eqref{eq:mph}. We reiterate that
only the deviatoric part of pressure Hessian needs to be modeled.
The architecture for viscous Laplacian
is essentially identical to that shown in Fig.~\ref{fig:tbnn}, with
the difference that the first output layer
and the second input layer have both 16 nodes, 
corresponding to the coefficients $c_2^{(i)}$ and $c_3^{(i)}$ 
and the tensors $\TT^{(i)}$ and $\BB^{(i)}$,
with appropriate prefactors corresponding to the Reynolds number scaling.

\section{Training and validation of ReS-TBNN model}
%DNS data, Training of the ReS-TBNN model and comparisons with DNS at higher $\re$

\subsection{DNS data}
To train the ReS-TBNN model, 
the `ground truth' data are obtained from 
a massive DNS database corresponding to 
forced stationary isotropic turbulence in a periodic domain
\cite{Ishihara09}.
The simulations were performed using Fourier pseudo-spectral 
methods \cite{Rogallo}, allowing us to obtain the data with 
the highest accuracy practicable. 
A key aspect of our data is 
that we have simultaneously achieved 
a wide range of Reynolds 
numbers and the necessary small-scale resolution to accurately resolve 
extreme events  \cite{BPB2020, BS2020}. 
Both of these conditions are indispensable for
successful model development. 
The Taylor-scale based Reynolds number $\re$ 
of our database ranges from $140-1300$. 
The data have been since utilized
and validated in several recent studies 
\cite{BP2021, BPB2022, BS2022, BS_PRL_2022}. 
A brief account of DNS and the database is provided in the Methods section, 
and more details can be found in the references just mentioned.
In order to train our network, only the data for 
$\re=140-650$ are utilized; subsequently, we will demonstrate that the trained network can successfully
predict statistics at higher (unseen) $\re=1300$. 
Though this $\re$ is only twice as large as the largest one used in the training, 
its usefulness should be assessed in the context of the computational expense of DNS, 
which would be easily 100 times more.
This is because the cost of DNS increases at least as 
strongly as $R_\lambda^6$, going up to $R_\lambda^8$ in the limit of large $\re$, 
to accurately resolve the smallest scales \cite{YS:05, BP2022}.

We point out that the model can also be trained over the range 
$\re=140-390$ and used to predict results at 
$\re=650$ and $1300$. However, learning from this smaller range of $\re$ is 
sub-optimal for capturing trends that can be extrapolated
to significantly higher $\re$. 
%For instance, $\re=140-390$
%gives a factor of about $390/140 \approx 2.8$ for the 
%learning range,
%whereas predicting at $\re=1300$ corresponds
%to extrapolating at a factor $1300/390 \approx 3.3$,
%and cannot be expected to work optimally. 
Additionally, learning from low $\re$ alone is not helpful because many 
features of turbulence are not fully developed for those conditions. 
For the present, learning from the range 
$\re=140-650$ and predicting the performance at $\re = 1300$ allows an
optimal situation. As remarked in the previous paragraph, 
the corresponding DNS effort would be much more expensive. Clearly, 
prediction over a wider range would be desirable. 
One can also use the full range of available of $\re=140-1300$ for learning,
and predicting at a higher unseen $\re$.
Since the DNS data at higher $\re$ are not yet available, 
the predictions would be unverifiable, so we leave this task for the future.

\subsection{Training of the ReS-TBNN model}

The training of the ReS-TBNN model is implemented 
in FORTRAN using a massively parallel
in-house deep-learning library.
We utilize the distributed training paradigm
with data parallelism, i.e.,
the training data is split across many processors,
with each processor having access to the same model.
The model parameters are synchronized via
inter-processor communication
after each training epoch
(executed using MPI collective communication calls). 
To update the parameters of the neural network, 
the quadratic loss function is minimized. For example,
for the pressure Hessian tensor, the loss function
is given by
\begin{align}
\mathcal{L} = \frac{1}{2 N_{\rm data}} \sum_{m=1}^{N_{\rm data}} 
|| \hat{\HH}_{\rm d}^{(m)} - \HH_{\rm d}^{(m)} ||_F^2.
\end{align}
where $\hat{\HH}_d$ is the model output
and $|| \cdot ||_F$ denotes the 
Frobenius norm. A similar loss function
can also be written for the viscous Laplacian term.

\begin{figure*}[ht]
\begin{center}
\includegraphics[width=0.88\textwidth]{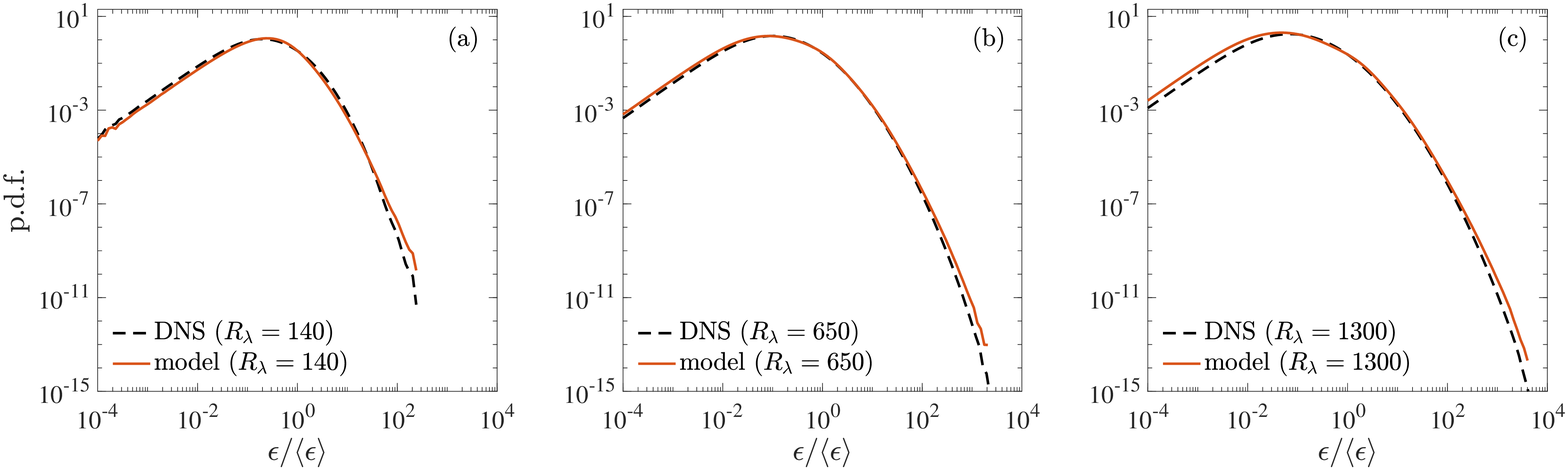} \\
\vspace{0.4cm}
\includegraphics[width=0.88\textwidth]{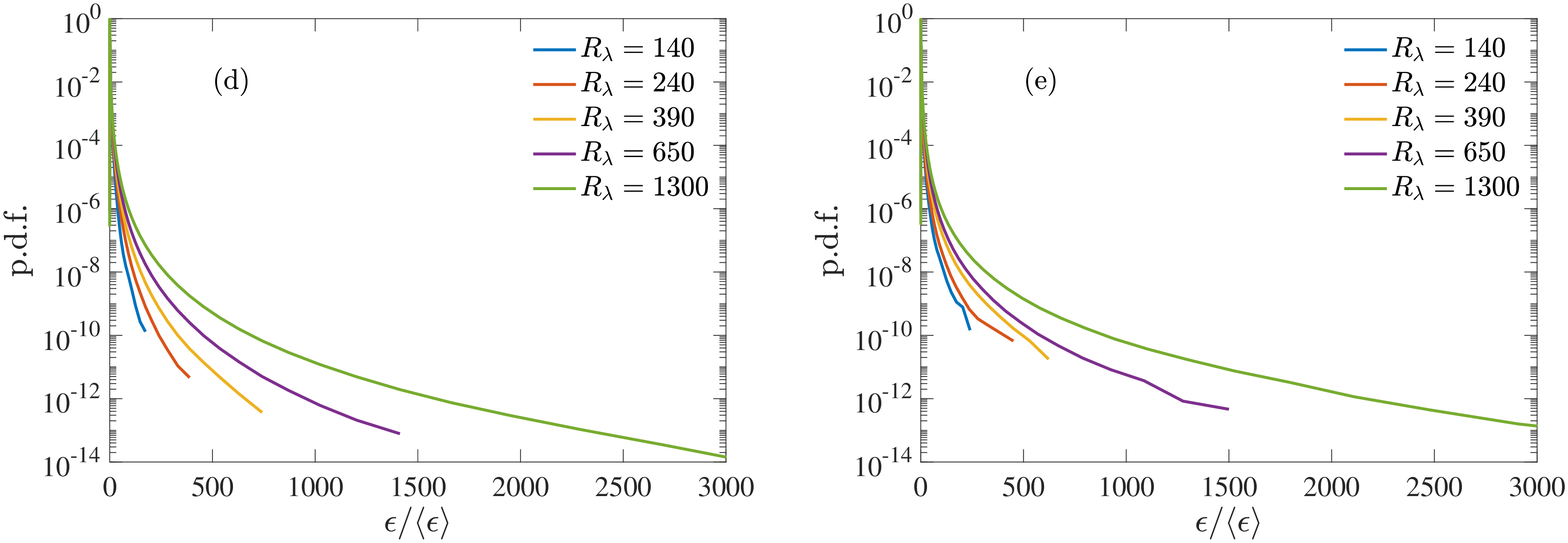}
\caption{
Comparisons of probability density functions (PDFs)
of the energy dissipation rate, non-dimensionalized
by the mean value, as obtained from network model
and DNS. Panels a-c show the comparison
at $\re=140$, $650$ and $1300$, respectively,
on log-log scales. 
Panel d shows the PDFs from DNS for 
various $\re$ on lin-log scales
highlighting the intermittency of PDF tails.
Panel e shows the PDFs obtained from
the network model for the same set of $\re$ as panel d,
showing the effectiveness of the model in predicting
the PDF tails. 
We reiterate that $\re=1300$ is never seen by the model.
}
\label{fig:diss}
\end{center}
\end{figure*}

The training data are compiled from 
DNS runs corresponding to $\re=140-650$. 
About one billion data points
are utilized for training, split evenly across
all $\re$ in the range. 
The training is performed on about a thousand processors
(with each processor handling one million data points).
Note that each data point corresponds to a combination
of the tensors $\AA$, $\HH_{\rm d}$, $\nabla^2 \AA$ 
and the particular value of $\re$. 
As discussed earlier, all variables are
non-dimensionalized by the Kolmogorov
scales.  The training is performed for many thousands of
epochs, until the loss function reaches a plateau
(see the Methods section). 
The learning rate is always kept low at $10^{-6}$.
As one might expect, the choice of the hyper parameters, such as the number
of hidden layers and the number of nodes per layer, 
plays a crucial role in obtaining the best
model. To select the optimal network configuration, we
do not utilize a validation data set, but
directly compare the velocity gradient statistics 
with DNS results. 
We found that a network with about 25 layers, each with about 50 nodes, 
provides optimal results; increasing the number of layers and
the number of nodes per layer results in only marginal improvement
(and can also lead to overfitting \cite{good2016}).

\section{Comparison of the ReS-TBNN model with DNS}

The effectiveness of the trained ReS-TBNN model 
will now be evaluated by
comparing its outcome with 
DNS results. We first focus
on the Reynolds number
trend of velocity gradient statistics (this being a key contribution of the model).
We particularly consider the probability density functions (PDFs) that display increasingly non-Gaussian tails with increasing Reynolds number, because of intermittency. 
All components of $\AA$ exhibit
intermittency, but it is convenient
to consider scalar quantities of 
direct physical significance, such as the energy dissipation
rate, whose mean value is the net energy flux
from large to small scales. As is well known, 
the instantaneous energy transfers are highly intermittent,
leading to extreme dissipation events \cite{Sreeni88}.

Figure~\ref{fig:diss} shows 
comparisons of the PDFs of the energy
dissipation rate, normalized by its mean value,
from DNS and the model. 
Panels a-c illustrate the comparison on log-log scales
at $\re=140$, $650$ and $1300$, respectively,
showing excellent agreement between the two results.
The ReS-TBNN model has been
trained only up to $\re=650$ and has not seen any data
for $\re=1300$.
For a closer inspection, 
panels d-e show the same comparisons on linear-log scales
for all $\re$ available.
The model captures the intermittent tails 
qualitatively well, though the extreme events are
overpredicted (see below).  We believe that this overprediction occurs because 
of the Reynolds number scaling
of the tensor bases; essentially, the rescaling
serves to normalize the tensor and the extreme events
have slightly stronger influence on the weights and biases. 
Note that, similar to the dissipation rate, one can also consider
other scalar measures derived from $\AA$, 
such as enstrophy $\Omega= \omega_i \omega_i $, 
where $\omega_i = \epsilon_{ijk} A_{jk}$ is the vorticity vector 
(with $\epsilon_{ijk}$ being the Levi-Civita symbol). 
Although not shown here, the agreement observed for enstrophy is similar.

\begin{figure}[ht]
\begin{center}
\includegraphics[width=0.36\textwidth]{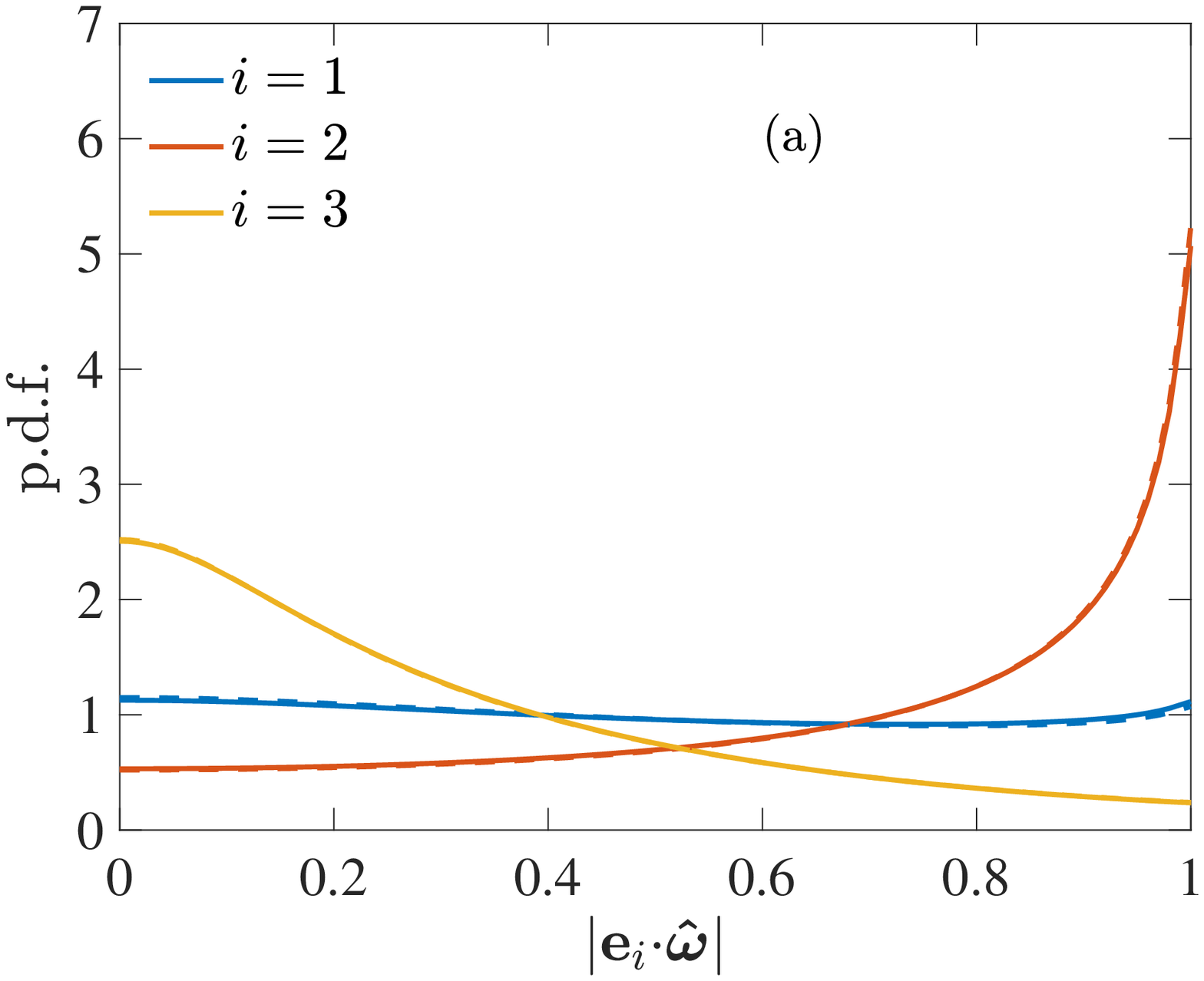}
\includegraphics[width=0.36\textwidth]{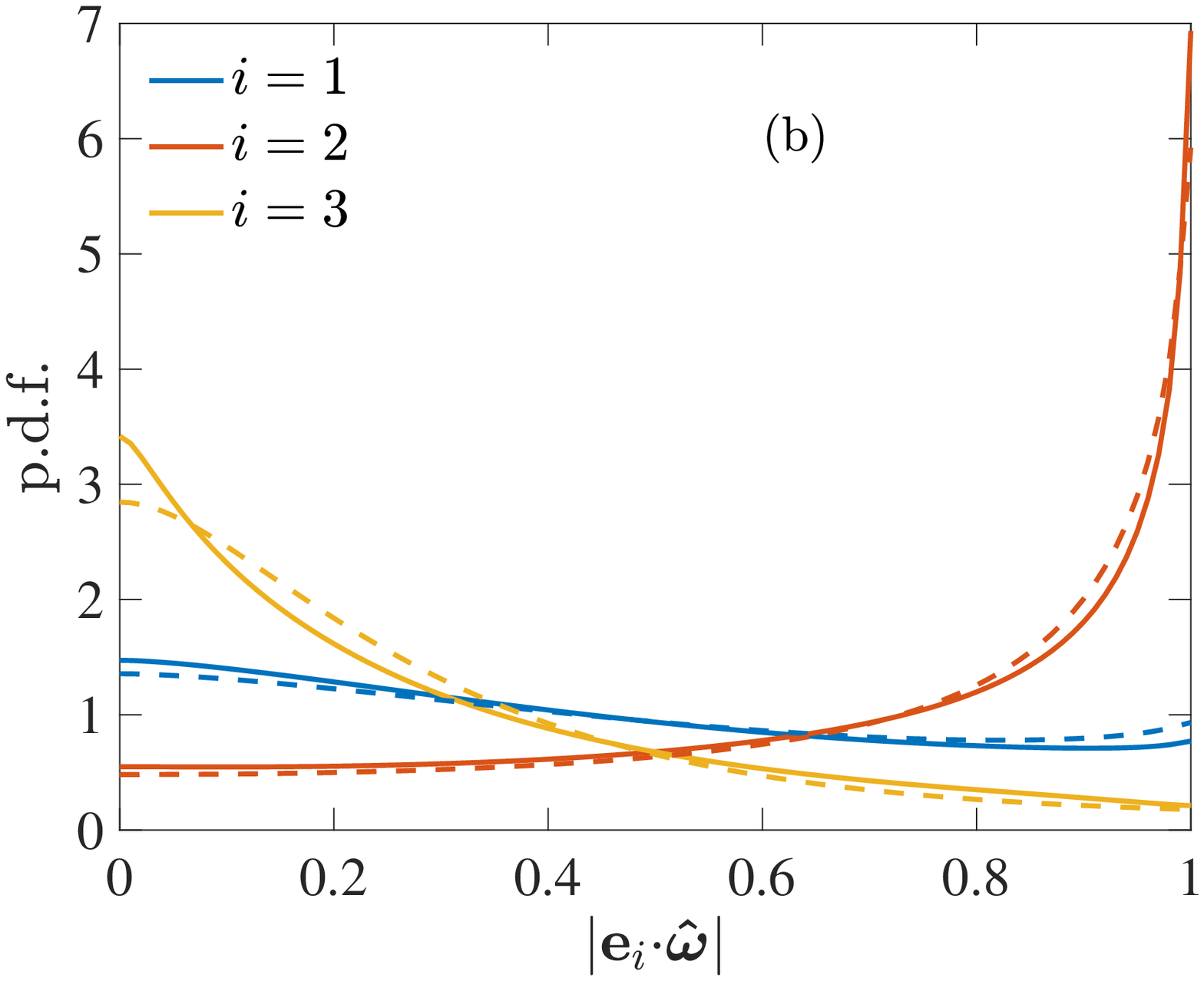}
\caption{
Comparison of the PDFs
of the cosine of the angles between the vorticity
unit vector $\hat{\boldsymbol{\omega}}$
and the eigenvectors of the strain tensor 
$\mathbf{e}_i$, corresponding to eigenvalues $\lambda_i$,
where $\lambda_1 \ge \lambda_2 \ge \lambda_3$. 
Panel a shows the result from DNS corresponding
to $\re=1300$ in solid lines and $\re=140$ in dashed lines.
Panel b shows the result from the network model corresponding
to same $\re$ values in solid and dashed lines.
The alignment PDFs have almost no $\re$ dependence
in DNS; that shown by the model is negligible also. 
}
\label{fig:align}
\end{center}
\end{figure}

\begin{table}
\begin{center}
    \begin{tabular}{l|c|c|c|c|c}
    \hline
    $\re$ & $140$ & $240$ & $390$ & $650$ & $1300$ \\
    \hline
    $\langle \epsilon^2 \rangle / \langle \epsilon \rangle^2 $  & ~ & ~ & ~ & ~ & ~  \\
    DNS & 2.75 & 3.28 & 3.86 & 4.67 & 6.27  \\
    model & 2.49 & 2.78 & 3.74 & 5.16 & 7.72  \\
    \hline
    $\langle \Omega^2 \rangle / \langle \Omega \rangle^2 $  & ~ & ~ & ~ & ~ & ~  \\
    DNS & 4.89 & 6.22 & 7.49 & 9.26 & 12.6  \\
    model & 4.48 & 5.50 & 7.25 & 9.80 & 14.4  \\   
    \hline
    $\langle \epsilon^4 \rangle / \langle \epsilon^2 \rangle^2 $  & ~ & ~ & ~ & ~ & ~  \\
    DNS & 37.8 & 83.8 & 171 & 387 &  1298  \\
    model & 101 & 260 & 434 & 628 & 2355  \\   
    \hline
    $\langle \Omega^4 \rangle / \langle \Omega^2 \rangle^2 $  & ~ & ~ & ~ & ~ & ~  \\
    DNS & 150 & 345 & 741 & 1636 & 6917  \\
    model & 316 & 812 & 1666 & 3101 & 13210  \\          
    \hline
    skewness $A_{11}$  & ~ & ~ & ~ & ~ & ~  \\
    DNS & -0.52 & -0.55 & -0.59 & -0.63 & -0.70  \\
    model & -0.49 & -0.54 & -0.60 & -0.66 & -0.75  \\
    \hline
    flatness $A_{11}$  & ~ & ~ & ~ & ~ & ~  \\
    DNS & 5.73 & 6.82 & 8.02 & 9.87 & 13.1  \\
    model & 5.20 & 5.80 & 7.80 & 10.9 & 15.2  \\
    \hline
    flatness $A_{12}$  & ~ & ~ & ~ & ~ & ~  \\
    DNS & 8.71 & 10.9 & 13.2 & 16.5 & 22.2  \\
    model & 8.07 & 9.90 & 13.0 & 17.8 & 25.8  \\
    \hline    
    \end{tabular}
\end{center}
\caption{
Comparison of various statistics 
from DNS and the model. 
Shown are the second and fourth order moments
of dissipation ($\epsilon$) and enstrophy ($\Omega$),
and the skewness and flatness factors of
longitudinal ($A_{11}$) and transverse ($A_{12}$) 
velocity gradient components. 
Note that the skewness of $A_{12}$ is not shown,
since it is zero (within statistical uncertainty), from
both DNS and the model.
}
\label{tab:red}
\end{table}

The over-prediction by the model occurs principally for events with probability 
less than about $10^{-9}$. Such events are obviously very important for high order 
moments, but the reliability of such very large moments is not quite assured 
for the DNS data itself. For example, suppose we compute the sixth moment of 
the energy dissipation rate. 
This is equivalent
to obtaining the $12$-th order moment of velocity gradients, 
which would be stretching one's credulity even for the large size of the present 
database. 
To better understand how well the model and the DNS agree, 
we directly compare some moments from the PDFs.
In Table~\ref{tab:red}, we list the second and fourth order moments
of dissipation and enstrophy from both the model and the DNS. 
We also compare the third and fourth
order moments of individual components of $\AA$. 
The results for skewness of $A_{12}$ are not
shown since it is zero (within statistical error)
for both the DNS and the model. 
Clearly the results from the model are very satisfactory 
for the second order of dissipation and enstrophy 
but less so for the fourth.

It is worth noting that the second order moments of 
dissipation and enstrophy can be directly related to
fourth order moments of velocity gradient components
in the following manner \cite{siggia:1981b}:
\begin{align}
\langle \epsilon^2 \rangle /\langle \epsilon\rangle^2 = \frac{15}{7} F(A_{11}) \ , \ \ 
\langle \Omega^2 \rangle /\langle \Omega\rangle^2 = \frac{9}{5} F(A_{12}) \ .   
\end{align}
Here, $F(A_{11})$ and $F(A_{12})$ are the flatness factors of
the components $A_{11}$ and $A_{12}$, respectively. 
These results are exact for isotropic turbulence at any Reynolds
number, and are nominally satisfied far
from solid boundaries in other turbulent flows
at high Reynolds numbers, i.e., when local isotropy holds
\cite{K41a, Frisch95}. 
It can be seen from Table~\ref{tab:red} that both these
relations are well satisfied in the model and DNS results.
In fact, many such isotropic relations exist
for various moment orders of velocity gradients.
For instance, for the second order moments, they are
\begin{align}
\begin{aligned}
&\langle A_{\alpha \alpha }^2 \rangle = \langle A_{\beta \beta }^2 \rangle \ , 
\ \ \  &\text{for} \ \alpha = \beta \\
&\langle A_{\alpha \beta }^2 \rangle =  2 \langle A_{\alpha \alpha }^2 \rangle \ , 
\ \ \  &\text{for} \ \alpha \ne \beta \\
&\langle A_{\alpha \alpha } A_{\beta \beta } \rangle = 
\langle A_{\alpha \beta } A_{\beta \alpha } \rangle = 
 - \langle A_{\alpha \alpha }^2 \rangle/2  \  
\ \ \  &\text{for} \ \alpha \ne \beta,
\end{aligned}
\end{align}
where repeated indices in $\alpha$ and $\beta$ do not imply summation.
Essentially, all second order moments can be simply
described by the moment $\langle A_{11}^2 \rangle $.
From the above, it also follows
\begin{align}
\begin{aligned}
\langle \epsilon \rangle/\nu &= 2 \langle S_{ij} S_{ij} \rangle 
= 15 \langle A_{11}^2 \rangle \ , \\
\langle \Omega \rangle &= 2 \langle R_{ij} R_{ij} \rangle 
= 15 \langle A_{11}^2 \rangle \ . 
\end{aligned}
\end{align}
Although not shown explicitly, we note that the above
relations are all satisfied in our model
results (and, of course, DNS) at all Reynolds numbers.

\begin{figure}[ht]
\begin{center}
\includegraphics[width=0.48\textwidth]{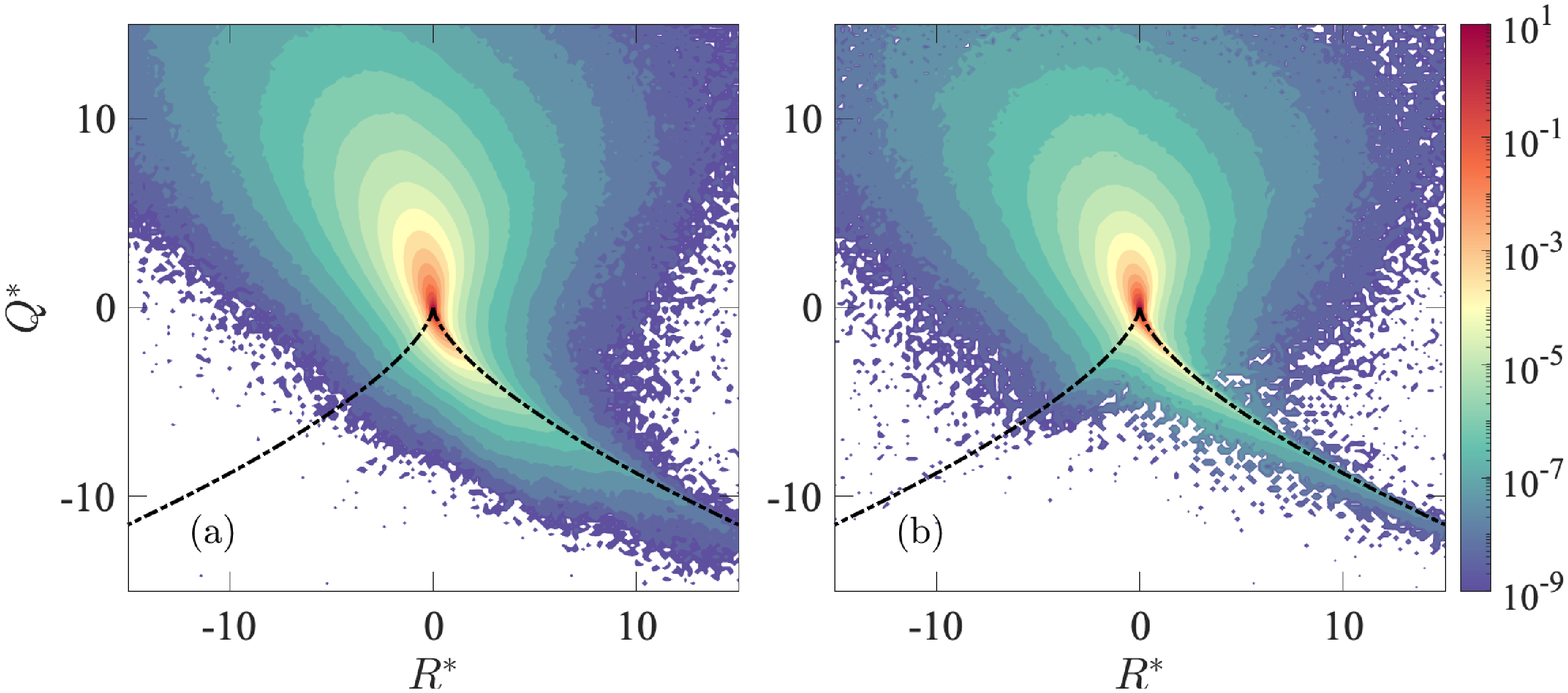}
\includegraphics[width=0.48\textwidth]{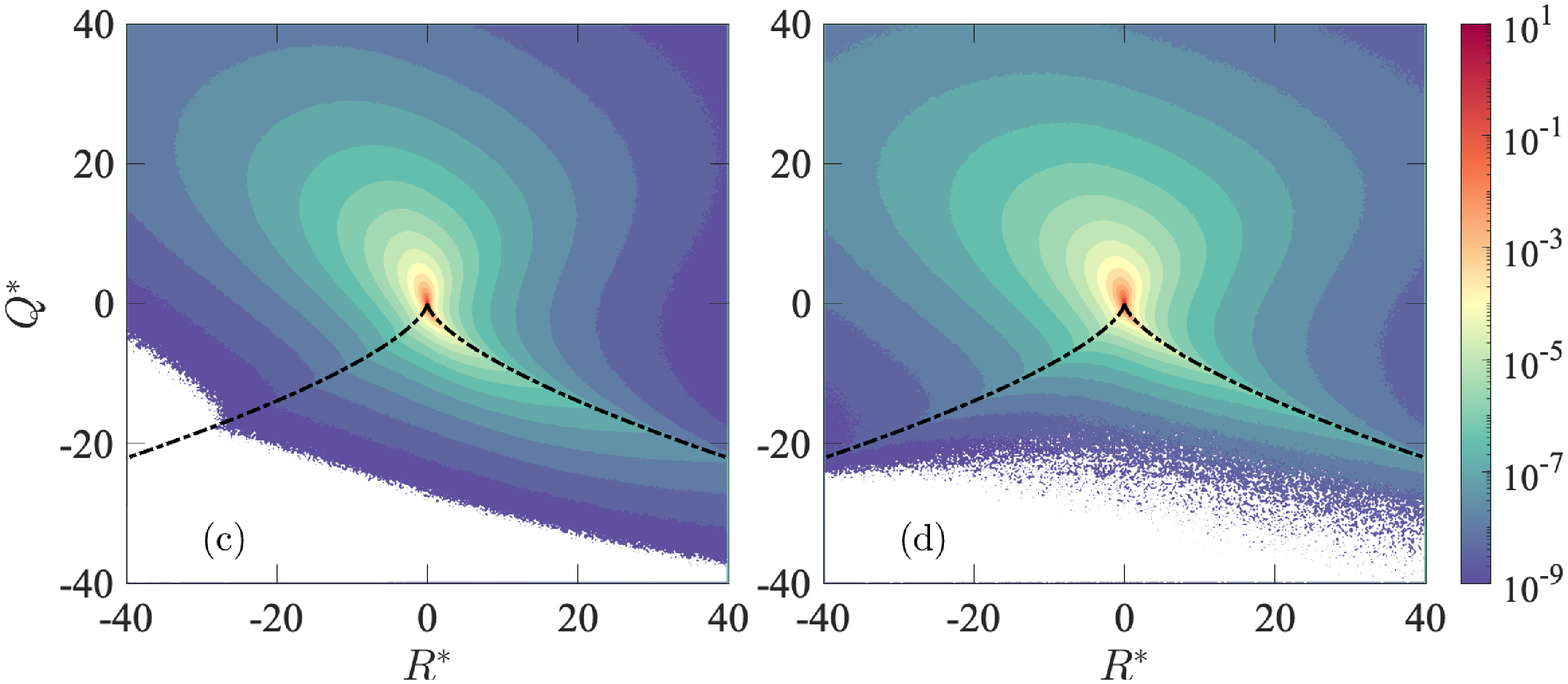}
\caption{
Comparison of joint probability density functions
of the invariants of the velocity gradient tensor,
defined as
$Q^* = -\frac{1}{2}\tr(\AA^2) \tau_K^2 $
and $R^* = -\frac{1}{3}\tr(\AA^3) \tau_K^3$.
Panel a shows the result
from DNS at $\re=140$ and
panel b shows the corresponding result
from the network model.
Panel c shows the DNS result 
at $\re=1300$ and 
panel d shows the corresponding result
from the model.
}
\label{fig:qr}
\end{center}
\end{figure}

The comparisons in Fig.~\ref{fig:diss} and Table~\ref{tab:red} 
predominantly focused on Reynolds number scaling
of various individual statistics. However, it is equally important
to capture the structure of velocity gradient
tensor. We examine two
well known universal results to this end, 
the first being the alignment of vorticity
vector with the eigenvectors of strain tensor, shown in Fig.~\ref{fig:align}.
Panel a shows the PDFs of the cosines of the alignment from DNS. 
Consistent with the well known result from the literature
\cite{Ashurst87}, vorticity preferentially
aligns with the second eigenvector of strain,
and is weakly orthogonal to third eigenvector;
whereas there is no preferential alignment with the
first eigenvector. There is virtually no
$\re$-dependence of these PDFs
(as noted in \cite{BBP2020}). 
In Fig.~\ref{fig:align}b the corresponding result is shown
from the ReS-TBNN model. 
The model captures the trends very well,
with slight enhancement of the respective alignments.
This trend is consistent with the result in 
Fig.~\ref{fig:diss} where the model
slightly overpredicts extreme events
(note that the alignments are
enhanced when considering extreme events \cite{BBP2020}).
We also note that the model shows only very weak Reynolds number dependence 
of the alignment properties, which is inconsequential 
for all practical purposes. 

The second structural aspect concerns the local 
critical point analysis of $\AA$,
which identifies the flow topology
using the second and third invariants of $\AA$ \cite{perry87}:
$Q = -\tr(\AA^2)/2$ 
and $R = -\tr(\AA^3)/3$. (Note that the first
invariant of $\AA$, i.e., its trace, is zero from incompressibility.)
The joint PDF of these two invariants is known to exhibit a universal
tear-drop shape \cite{Tsi2009, Meneveau11}. 
For a final comparison between
DNS and model, we compare the joint PDF in  Fig.~\ref{fig:qr}a-b 
obtained from DNS and the model, respectively, at $\re=140$;
the same results for $\re=1300$ are shown in Fig.~\ref{fig:qr}c-d. 
In both cases, the model
predicts the joint PDF quite well in all
four quadrants. It is worth noting that
the joint PDFs also exhibit strong
intermittency with increasing Reynolds number;
this aspect is again well captured by the
model, similar to the result in Fig.~\ref{fig:diss}.

As a final remark, we note that 
one can compare many other quantities
to evaluate the performance of the model
(in relation to DNS). For instance, 
one can consider the dynamics of velocity gradients
projected onto the $Q-R$ plane \cite{Wilczek2014Pressure, Johnson2016Closure, tian21}, 
or how they influence trajectories of 
particles in turbulence \cite{Leppin2020Capturing}. 
Such comprehensive studies, including detailed comparisons
of our model with the previous
ones, will be presented in a subsequent
paper.
In the current study, we have highlighted
the most important contribution of our model, which
is to capture intermittency and Reynolds numbers trends
of velocity gradient statistics.

\section{Discussion}

In turbulence, the DNS of Navier-Stokes equations 
on massive supercomputers
is now an established area for gaining a fuller understanding of flow physics, 
leading to more reliable predictions. 
However, both theoretical and practical needs demand ever-increasing size of computations, so that fluid turbulence will always remain, for the foreseeable future, 
as one of the frontier computational problems, no matter how large the supercomputers become. 

In this regard, the major bottleneck is the need to simulate small scales of turbulence with
high fidelity (adequate resolution in space and time, convergence, etc). 
To make progress on real problems, one needs to model small scales well,
for instance in large-eddy simulations (LES) where large scales are resolved, 
but small scales are modeled
assuming a degree of universality. 
This modeling approach has been largely guided by `human
learning', often resulting in {\em ad hoc} considerations 
depending on the flow.
%This approach has led to various turbulence models,
%largely guided by `human learning', 
%with various unknown constants and coefficients that are obtained empirically 
%from experiments or DNS by comparing various average quantities of interest; 
As modern methods of deep machine learning have expanded,
it appears possible for them to aid in modeling by directly 
learning from vast amount of high-fidelity data that is already
available over some range of Reynolds numbers. 
In this scenario, deep neural networks are allowed to do the fitting 
at a deeper level of  instantaneous data,
in the process satisfying a substantially larger set of constraints than
possible by `human learning'. 
If this attempt succeeds, we will have a powerful tool in assimilating the 
lower Reynolds number data for predicting flow properties at higher
unseen Reynolds numbers. This is a difficult problem given the nature of turbulence.

In this paper, we have made a ground-level attempt towards our stated goal. 
We have demonstrated that the small scale dynamics of turbulence,
as captured by velocity gradients,
can be modeled quite well using deep
neural networks. We have trained the network
on a range of Reynolds numbers available 
from DNS, and leveraged this training to predict
results at higher Reynolds numbers, whose properties the network 
does not know in advance. 
The effort is encouraging 
not only in predicting the intermittency of velocity gradients
with increasing Reynolds number,
but also various signature topological properties of 
velocity gradient tensor, such as alignment of vorticity with
strain rate eigenvectors, and the joint PDFs of invariants displaying
a tear-drop shape.

There are certain shortcomings of the trained model
when considering truly extreme events, which contribute
significantly only to higher order moments. 
It might be possible to further improve
this aspect by incorporating the current deep learning
approach in alternative frameworks for velocity 
gradient dynamics,
which incorporate Reynolds number dependencies more
naturally, see e.g. Refs.~\cite{Johnson2017Turbulence, das2019}.

Finally, it would also be worth  
expanding the current effort in a more
concerted way to
other modeling paradigms such 
as LES \cite{beck2019} -- 
allowing one to tackle more complex
turbulent flows at Reynolds numbers of practical
interest in nature and engineering.
Such an extension can be readily managed, for instance,
by considering filtered velocity gradient tensor,
which would be amenable to the same tensor framework 
as utilized here \cite{ling2016, tian21}.
Likewise, the framework
developed here can also be extended to 
study the dynamics of scalar gradients
in turbulent mixing problems \cite{buaria_dfd22}, 
especially in the high Schmidt number regime;
though these conditions are even more challenging
for DNS \cite{yeung2005}, recent efforts have led to 
generation of high fidelity data
at reasonably high Reynolds numbers
\cite{BCSY2021a, BCSY2021b}. 
Efforts in these directions are under way 
and will be reported as future work.

\section{Methods}

\subsection{Direct numerical simulations}
The data utilized here are obtained by the
DNS of incompressible Navier-Stokes
equations
\begin{align}
\partial \uu/\partial t + \uu \cdot \nabla \uu =
-\nabla P  + \nu \nabla^2 \uu + \mathbf{f} \ 
\label{eq:ns}
\end{align}
where $\uu$ is the velocity, satisfying $\nabla \cdot \uu=0$,
$P$ is the kinematic pressure and $\nu$ is the kinematic
viscosity. The term $\mathbf{f}$ corresponds to large-scale
forcing required to maintain a statistically stationary
state.  The simulations correspond to the canonical setup of
isotropic turbulence with periodic boundary conditions 
in a cubic domain of side length $L_0 = 2\pi$.
It is well known that such a setup allows one to
reach the highest Reynolds numbers in DNS and is ideal 
for studying small scales \cite{Ishihara09}. 
Taking the gradient of Eq.~\eqref{eq:ns}
leads to Eq.~\eqref{eq:dadt}, without the forcing term. 
When performing the Monte-Carlo simulations
of ReS-TBNN model, a forcing term is reintroduced
to mimic this effect and achieve stationary statistics
(as described in the subsection after the next). 

\begin{table}[tbhp]
\centering
\caption{Various simulation parameters for the DNS runs
utilized here; 
the Taylor-scale Reynolds number ($\re$),
the number of grid points ($N^3$),
spatial resolution ($k_{\rm max}\eta$), 
ratio of large-eddy turnover time ($T_E$)
to Kolmogorov time scale ($\tau_K$),
length of simulation ($T_{\rm sim}$) in statistically stationary state
}
    \begin{tabular}{ccccc}
    $\re$   & $N^3$    & $k_{\rm max}\eta$ & $T_E/\tau_K$ & $T_{\rm sim}$   \\
\hline
    140 & $1024^3$ & 5.82 & 16.0 & 6.5$T_E$  \\
    240 & $2048^3$ & 5.70 & 30.3 & 6.0$T_E$  \\
    390 & $4096^3$ & 5.81 & 48.4 & 4.0$T_E$  \\
    650 & $8192^3$ & 5.65 & 74.4 & 2.0$T_E$  \\
   1300 & $12288^3$ & 2.95 & 147.4 & 20$\tau_K$  \\
%    ~   & ~    & ~ & ~ & ~ & ~ & ~ \\
\hline
    \end{tabular}
\label{tab:dns}
\end{table}

The DNS domain consists of $N^3$ grid points
with uniform grid spacing $\Delta x = L_0/N$ in each direction.
The equations are solved using a massively parallelized version of the 
well-known Fourier pseudospectral algorithm of Rogallo \cite{Rogallo};
the resulting aliasing errors are controlled by a combination of grid shifting
and spherical truncation \cite{PattOrs71}. For time integration,
second-order Runge-Kutta method is used, with the time step
$\Delta t$ subject to the Courant number $C$ constraint 
for numerical stability: $\Delta t = C \Delta x /||\uu||_{\infty}$
(where $||\cdot||_{\infty}$ is the $L^\infty$ norm).
An important consideration in studying velocity gradients
and associated extreme events is that of spatial resolution, 
captured by the ratio $\Delta x /\eta_K$ ($\eta_K$ being
the Kolmogorov length scale, defined earlier in Eq.~\eqref{eq:Kol}).
For pseudospectral DNS, spatial resolution is also prescribed by
the parameter $k_{\rm max}\eta_K$, where $k_{\rm max} = \sqrt{2}N/3$
is the maximum resolved wavenumber. It can be easily shown that
$\Delta x/ \eta \approx 3/k_{\rm max}\eta_K$. 
All our runs correspond to high spatial resolution,
going up to $k_{\rm max}\eta_K \approx 6 $ to accurately resolve 
extreme events. 
The DNS database, along with various simulation parameters, are
summarized in Table~\ref{tab:dns}.

We note that one can also utilize Lagrangian 
data, as obtained from 
following fluid particle trajectories
along with the Eulerian DNS \cite{yeung2006},
to train and validate the 
deep learning network \cite{tian21}. 
However, since the network
relies on obtaining a local functional
closure, it does not make a difference
whether Eulerian or Lagrangian data are utilized,
provided they both are statistically
stationary. Note that  Lagrangian data
are obtained from Eulerian data using spline interpolation
\cite{PopeYeung88, buaria.cpc},
and thus are less accurate, especially
for higher order moments \cite{yeung2006, BS_PRL_2022}.
Even with this caveat, it would be desirable to 
construct recurrent neural networks \cite{good2016}, 
to enable direct learning of
both spatial and 
temporal dependencies in the data,
possibly leading to improved predictive
capabilities.

\subsection{ReS-TBNN loss function}

\begin{figure}
\begin{center}
\includegraphics[width=0.45\textwidth]{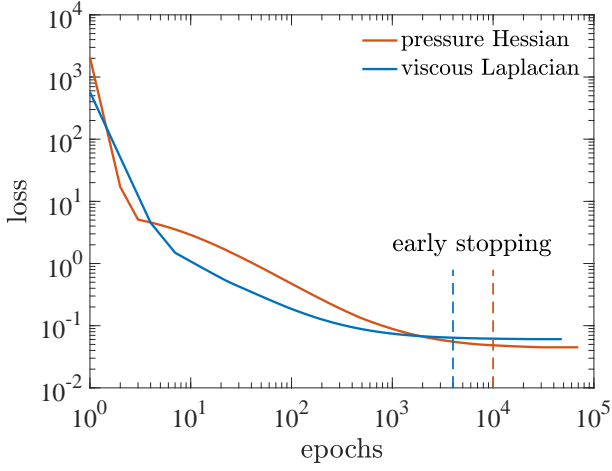}
\caption{
Decay of the loss function during training 
of the ReS-TBNN for pressure Hessian
and viscous Laplacian terms. 
One epoch corresponds to the entire set of 
available training data spanning $\re=140-650$.
}
\label{fig:loss}
\end{center}
\end{figure}

Figure~\ref{fig:loss} shows the behavior
of the loss functions, for training of both pressure Hessian
and viscous Laplacian terms, versus the number of epochs
lapsed as the training is performed.
Evidently, they 
become flat beyond 
a certain point.
An early stopping criterion,
as marked in Fig.~\ref{fig:loss},
is utilized for training 
of the networks
to avoid overfitting
\cite{girosi1995}.

\subsection{Monte-Carlo simulations of Res-TBNN model}

Once the pressure Hessian and viscous Laplacian
terms are modeled as functions of $\AA$, we can obtain
a closed system in $\AA$, which can be solved for arbitrary 
initial conditions. However, it is also necessary to add
a forcing term to the model \cite{girimaji1990}, 
mimicking the effects of large-scale forcing, to 
achieve stationary statistics.
Additionally, the forcing term also acts to reproduce
some effects of nonlocality (of pressure Hessian), which are lost
due to a local functional closure 
\cite{chevillard06prl, Wilczek2014Pressure, Johnson2016Closure}.
Thus, the closed system is posed in the form of a stochastic ODE, 
given as 
\begin{align}
\begin{aligned}
d \AA^* = \mathcal{F}(\AA^*) dt^* + d\FF^* .
\end{aligned}
\end{align}
Here, $\mathcal{F}(\AA^*)$ is the deterministic
tensor function, 
obtained from Eqs.~\eqref{eq:dadt2}-\eqref{eq:mvl} 
and given as
\begin{align}
\begin{aligned}
\mathcal{F}(\AA^*) = &- ({\AA^*}^2 - \frac{1}{3} \tr({\AA^*}^2) \II ) 
- \sum_{i=1}^{10} c_1^{(i)} R_\lambda^{-\beta_1^{(i)}} \TT^{*(i)}   \\
&+  \sum_{i=1}^{10} c_2^{(i)} R_\lambda^{-\beta_2^{(i)}} \TT^{*(i)}  
+ \sum_{i=1}^{6} c_3^{(i)} R_\lambda^{-\beta_3^{(i)}} \BB^{*(i)}  
\end{aligned}
\label{eq:dadt3}
\end{align}
and $d\FF^*$ is the stochastic forcing term
\begin{align}
dF_{ij}^* = b_{ijkl} dW_{kl}
\end{align}
built on tensorial Wiener process,
i.e., $\langle dW_{ij} \rangle=0$
and $\langle dW_{ij} dW_{kl} \rangle= \delta_{ik}\delta_{jl} dt^*$
with diffusion tensor $D_{ijkl} = b_{ijmn} b_{klmn}$.

For the drift term $b_{ijkl}$, we utilize the 
result of \cite{Johnson2016Closure}
\begin{align}
\begin{aligned}
b_{ijkl} = - \frac{1}{3} D_S \delta_{ij} \delta_{kl} 
+ \frac{1}{2} \left( D_S + D_R \right) \delta_{ik} \delta_{jl} \\
+ \frac{1}{2} \left( D_S - D_R \right) \delta_{il} \delta_{jk} \\
\end{aligned}
\end{align}
where $D_S$ and $D_R$ are free parameters that can tuned
to appropriately force the symmetric and skew-symmetric
parts of $\AA^*$, i.e., $\SS^*$ and $\RR^*$,
respectively, allowing us to impose consistency conditions
for stationarity:
$\langle S^*_{ij} S^*_{ij} \rangle = 1/2 $
and $\langle R^*_{ij} R^*_{ij} \rangle = 1/2 $.
Note that the former follows directly from the definition
of Kolmogorov time scale, whereas the latter
follows from statistical homogeneity. 
It also readily follows 
that $\langle A^*_{ij} A^*_{ij} \rangle = 1$.
The Monte-Carlo simulations are performed
starting from random Gaussian initial conditions of $\AA^*$,
until a stationary state as prescribed by
above conditions is reached. Thereafter, the simulations
are extended for desired duration to obtain converged
statistics.  

We note that given the complex functional form of the
deep learning closure, we encounter some rogue
trajectories (since the deep learning based closure does not 
guarantee stability). The encounter rate is only about one in a million.
Such trajectories are simply discarded from the ensemble 
but, if desired, can be regularized as
described in \cite{Leppin2020Capturing}.

\section*{Acknowledgements}

We gratefully acknowledge the Gauss Centre for Supercomputing 
e.V. (www.gauss-centre.eu) for providing computing time on the 
supercomputers  JUQUEEN and JUWELS at J\"ulich Supercomputing 
Centre (JSC), where the simulations and analyses reported in this
paper were primarily performed.

%\bibliography{large_grad}

%merlin.mbs apsrev4-1.bst 2010-07-25 4.21a (PWD, AO, DPC) hacked
%Control: key (0)
%Control: author (0) dotless jnrlst
%Control: editor formatted (1) identically to author
%Control: production of article title (0) allowed
%Control: page (1) range
%Control: year (0) verbatim
%Control: production of eprint (0) enabled
%

\end{document}